\documentclass{amsart}
\usepackage{amssymb}
\usepackage{amsthm}
\usepackage{amstext}
\usepackage{amsbsy}
\usepackage{amscd}
\usepackage{enumerate}
\usepackage{subfigure}
\usepackage[noadjust]{cite}
\usepackage[linktocpage]{hyperref}
\usepackage{graphicx}
\usepackage{mathtools}
\usepackage{pict2e}
\usepackage[usenames,dvipsnames]{color}
\usepackage[toc,page]{appendix}
\usepackage{tikz}
\usepackage{verbatim}
\usetikzlibrary{matrix}

\newtheorem{thm}{Theorem}[section]

\theoremstyle{remark}
\newtheorem{rem}[thm]{Remark}
\theoremstyle{definition}

\newcommand{\norm}[1]{\left\Vert#1\right\Vert}

\newcommand{\set}[1]{\left\{#1\right\}}
\newcommand{\R}{\mathbb{R}}

\newcommand{\N}{\mathbb{N}}
\newcommand{\Z}{\mathbb{Z}}

\newcommand{\eqdef}{\overset{\mathrm{def}}=}

\newcommand{\shd}{\mathrm{HD}}

\usepackage{array}
\newcolumntype{L}{>{\arraybackslash}m{6cm}}

\begin{document}

\title[HD of the spectrum of the square Fibonacci Hamiltonian]{Hausdorff dimension of the spectrum of the square Fibonacci Hamiltonian}

\author[W. Yessen]{William Yessen}
\email{yessen@rice.edu}
\address{Mathematics, Rice University, 1600 Main St. MS-136, Houston, TX 77005}

\thanks{\indent Supported by the NSF grant DMS-1304287.}


\date{\today}

\begin{abstract}

Denoting the Hausdorff dimension of the Fibonacci Hamiltonian with coupling $\lambda$ by $\shd_\lambda$, we prove that for all but countably many $\lambda$, the Hausdorff dimension of the spectrum of the square Fibonacci Hamiltonian with coupling $\lambda$ is $\min\set{2\shd_\lambda, 1}$. Our proof relies on the dynamics of the Fibonacci trace map in combination with the recent result of M. Hochman and P. Shmerkin on the Hausdorff dimension of sums of Cantor sets which are attractors of regular iterated function systems (\emph{Local entropy averages and projections of fractal measures}, Ann. Math. \textbf{175} (2012), 1001--1059).

\end{abstract}

\maketitle


\section{Introduction}
Recall that the Fibonacci Hamiltonian, a bounded self-adjoint operator acting on $\ell^2(\Z)$, is defined as
\begin{align*}
(H_\lambda\phi)_n = \phi_{n+1}+\phi_{n-1}+\lambda\omega_n\phi_n,
\end{align*}
with $\lambda > 0$ and $\omega=\set{\omega_n}_{n\in\Z}$ given by
\begin{align*}
\omega_n = \chi_{[1-\alpha, 1)}(n\alpha + \omega_0 \mod 1),
\end{align*}
where $\alpha = \frac{\sqrt{5}-1}{2}$, the inverse of the golden mean, and $\omega_0\in\R/\Z$. This operator has been widely studied in the context of electronic transport properties of quasicrystals for the past thirty years (see \cite{Damanik2013,Damanik2014e} and references therein for details). It is known that the spectrum of $H_{\lambda}$ is independent of $\omega_0$, and is a Cantor set of zero Lebesgue measure. Fractal properties of the spectrum, such as its Hausdorff dimension, play an important role in the understanding of the quantum dynamics \cite{Damanik2010,Last1996}. Today, the spectrum as well as the quantum dynamical properties of $H_\lambda$ are more or less completely understood \cite{Damanik2014e}. Recently, the focus has began to shift towards the so-called \emph{square Fibonacci Hamiltonian}, which acts on $\ell^2(\Z^2)$ and is given by
\begin{align*}
(H^2_\lambda\phi)_{n, m} = \phi_{n+1, m} + \phi_{n-1, m}
+ \phi_{n, m+1} + \phi_{n, m-1}
+ \lambda(\omega_n + \omega_m)\phi_{n, m};
\end{align*}
(see \cite{Damanik2013X} and references therein). Unlike in the one-dimensional case, very little is known about the square Fibonacci Hamiltonian.

Let us denote the spectrum of $H_\lambda$ by $\Sigma_\lambda$, and that of $H_\lambda^2$ by $\Sigma_\lambda^2$. It is known from the general principles in spectral theory (see, for example, Appendix A in \cite{Damanik2013X}) that
\begin{align*}
\Sigma_\lambda^2=\Sigma_\lambda + \Sigma_\lambda\eqdef\set{a + b: a, b\in\Sigma_\lambda}.
\end{align*}
Denote the Hausdorff dimension of $\Sigma_\lambda$ by $\shd_\lambda$, and of $\Sigma_\lambda^2$ by $\shd_\lambda^2$. Given that $\Sigma_\lambda$ is a Cantor set, questions about the topology and the fractal dimensions of $\Sigma_\lambda^2$ are highly nontrivial, while such detailed information about $\Sigma_\lambda^2$ is desirable. It is known that when $\lambda > 0$ is sufficiently small, $\Sigma_\lambda^2$ is an interval. It is also known that for all $\lambda$ sufficiently large, $\Sigma_\lambda^2$ is a Cantor set of Hausdorff dimension strictly smaller than one, and hence of zero Lebesgue measure. These results rely on quantitative estimates of $\shd_\lambda$ (see \cite{Damanik2013} for an overview), but do not give $\shd_\lambda^2$ explicitly in terms of $\shd_\lambda$. In general, however, it is known that
\begin{align}\label{eq:ineq}
\shd_\lambda^2\leq \min\set{2\shd_\lambda, 1}
\end{align}
(see \cite[Theorem 8.10(2)]{Mattila1995} and use the fact that the box-counting dimension of $\Sigma_\lambda$ coincides with its Hausdorff dimension -- see \cite[Theorem 1.1]{Damanik2014e}). In this paper we prove

\begin{thm}\label{thm:thm2}
For all by countably many $\lambda > 0$, we have
\begin{align*}
\shd_\lambda^2 = \min\set{2\shd_\lambda, 1}.
\end{align*}
\end{thm}

\begin{rem}
We believe the statement of Theorem \ref{thm:thm2} holds for all $\lambda$. Also, due to our techniques, the countable set of exceptions turns out to be dense in $(0,\infty)$.
\end{rem}

Let us point out that slightly more general families of square Hamiltonians have been considered (e.g. \cite{Damanik2013X} and references therein) where in the definition of $H_\lambda^2$, $\lambda(\omega_n + \omega_m)$ is replaced by $\lambda_1\omega_{1,n} + \lambda_2\omega_{2,m}$, $\lambda_1, \lambda_2\in(0,\infty)$, and for $i=1,2$, $\omega_{i,k}=\chi_{[1-\alpha, 1)}(k\alpha+\omega_i\mod 1)$ for some $\omega_i\in \R/\Z$. Let us denote this more general Hamiltonian by $H_{(\lambda_1, \lambda_2)}^2$, and the Hausdorff dimension of its spectrum by $\shd_{(\lambda_1, \lambda_2)}^2$. The spectrum of the Hamiltonian is given by $\Sigma_{\lambda_1}+\Sigma_{\lambda_2}$, where $\Sigma_{\lambda_i}$ is the spectrum of the Hamiltonian $H_{\lambda_i}$ (see Appendix A in \cite{Damanik2013X}). Our techniques apply to this case as well; that is, we have

\begin{thm}\label{thm:thm1}
For every $\lambda_1\in (0,1)$ fixed and for all but countably many $\lambda_2\in (0,\infty)$, we have
\begin{align*}
\shd_{(\lambda_1, \lambda_2)}^2 = \min\set{\shd_{\lambda_1}+\shd_{\lambda_2}, 1}.
\end{align*}
\end{thm}

Our proof relies on the dynamics of the Fibonacci trace map and the recent theorem of M. Hochman and P. Shmerkin on the Hausdorff dimension of sums of regular\footnote{Sometimes regular Cantor sets are also called \emph{dynamically defined}.} Cantor sets \cite{Hochman2012}. The Hochman-Shmerkin result has been applied in many works since it first appeared; however, to the best of our knowledge, the present work is the first application to a concrete physical model.

We should also point out that the Hochman-Shmerkin theorem is a generalization of previous results \cite{Moreira1998,Nazarov2012,Peres2009}; however, we had not been able to verify the hypothesis of the previous theorems (we emphasize in particular \cite{Moreira1998}\footnote{Via private communication, Moreira presented to us a sketch of a proof that a given pair of Cantor sets satisfies the hypothesis of the Hochman-Shmerkin theorem provided that one of them is an attractor of a $C^2$ iterated function system that is not conjugate to a linear one; however, we work with systems that are $C^{1+\alpha}$, $\alpha\in(0,1)$.}).

\section{Proof}

As mentioned above, our proof relies on the dynamics of the Fibonacci trace map and an application of the Hochman-Shmerkin theorem. Let us briefly describe our approach.

\subsection{Background}

For $E\in\R$, define
\begin{align}\label{eq:line}
\ell_\lambda(E) = \left(\frac{E-\lambda}{2}, \frac{E}{2}, 1\right).
\end{align}
Notice that $\ell_\lambda$ is a line lying on the smooth surface
\begin{align*}
S_\lambda\eqdef\set{(x,y,z)\in\R^3: x^2 + y^2 + z^2 - 2xyz-1 = \frac{\lambda^2}{4}}.
\end{align*}
Now define the \emph{Fibonacci trace map} $f: \R^3\rightarrow \R^3$ by
\begin{align*}
f(x,y,z)=(2xy-z, x, y).
\end{align*}
It is easily verified that $f(S_\lambda)=S_\lambda$ for every $\lambda$. Furthermore, for every $\lambda \neq 0$, $f|_{S_\lambda}$ is an Axiom A diffeomorphism \cite{Casdagli1986,Damanik2009,Cantat2009}. It is known that $E\in\Sigma_\lambda$ if and only if the forward orbit of $\ell_\lambda(E)$, $\set{f^n(\ell_\lambda(E))}_{n\in\N}$, is bounded \cite{Suto1987}. Using the Axiom A property of $f$, it is proved that $\ell_\lambda(\Sigma_\lambda)$ is precisely the intersection set of the line $\ell_\lambda(\R)$ with the stable lamination. This implies that $\Sigma_\lambda$ is a regular Cantor set \cite[Theorem 1.1]{Damanik2014e}. The conclusion of Theorem \ref{thm:thm2} is then obtained by an application of the Hochman-Shmerkin theorem (Corollary 1.5 in \cite{Hochman2012}). Thus, the proof of Theorem \ref{thm:thm2} consists of verifying the assumptions of the Hochman-Shmerkin theorem.

\subsection{Proof  of Theorem \ref{thm:thm2}}

It is known that the stable lamination on $S_\lambda$ intersects $\ell_\lambda$ transversally \cite[Section 2]{Damanik2014e}. Furthermore, it is easily deduced from \cite[Section 3]{Damanik2009}, using continuity of the stable lamination in the parameter $\lambda$, that for all $\lambda > 0$ sufficiently small and for every $f$-periodic point $p\in S_\lambda$, the stable manifold of $p$, $W^s(p)$, intersects $\ell_\lambda$. It follows that for all $\lambda > 0$ and every $f$-periodic $p\in S_\lambda$, $W^s(p)$ intersects $\ell_\lambda$.

\begin{centering}
\begin{figure}[t]
\includegraphics[scale=.8]{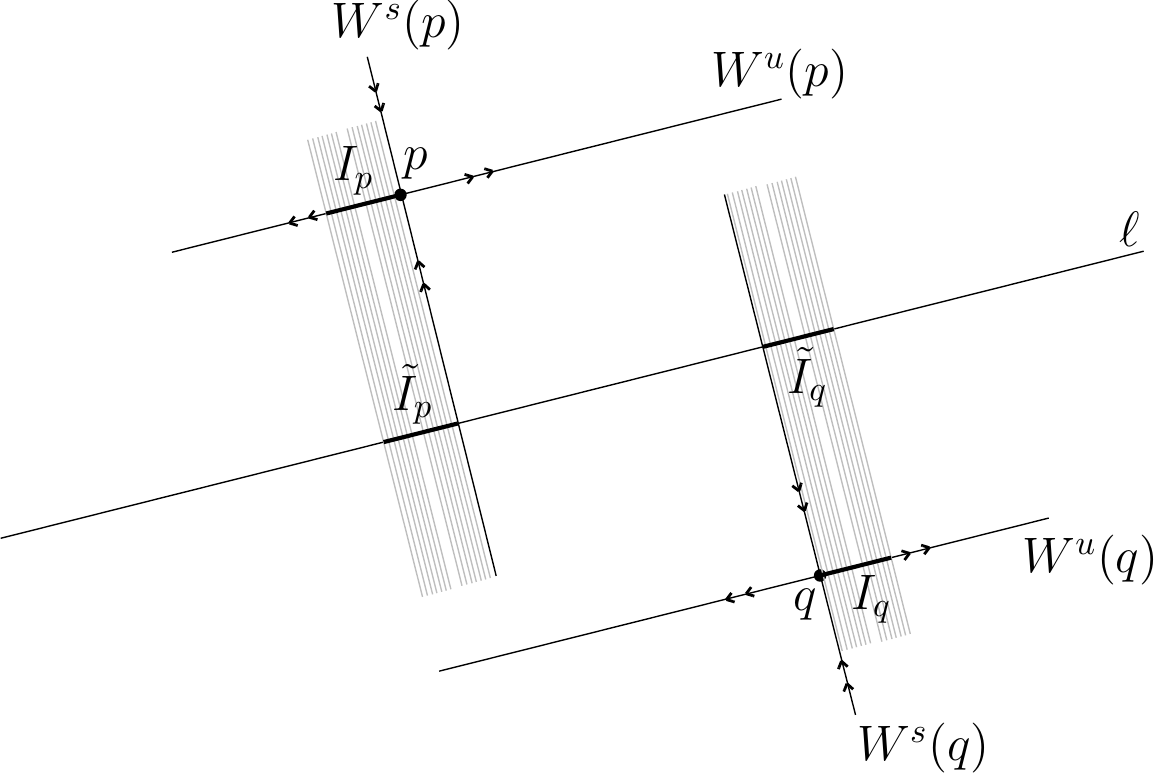}
\caption{}
\label{fig:holonomy}
\end{figure}
\end{centering}

Now, for any $\lambda > 0$, take two periodic points $p, q \in S_{\lambda}$ not in the same periodic orbit of (minimal) periods $n_p$ and $n_q$, respectively. Denote their stable manifolds by $W^s(p)$ and $W^s(q)$, and the unstable manifolds by $W^u(p)$ and $W^u(q)$, respectively. Denote the stable lamination on $S_{\lambda}$ by $W^s_\lambda$. Then for every $\epsilon > 0$ there exists a compact interval along $W^u(p)$ (respectively, $W^u(q)$) of length at most $\epsilon$ containing $p$ (respectively, $q$) as one of its endpoints, which we denote by $I_p$ (respectively, $I_q$) such that the intersection of this interval with $W^s_\lambda$ is a Cantor set; moreover, this Cantor set can be realized as the attractor of a regular Iterated Function System, or IFS for short (see \cite[\S 9 and \S 11]{Hochman2012} for definitions). One of the contracting functions of this IFS is $f^{-n_p}|_{W^u(p)}$ (respectively, $f^{-n_q}|_{W^u(q)}$) whose fixed point is $p$ (respectively, $q$) (see, for example, \cite[\S 1 in Chapter 4 and Appendix 2]{Palis1993} for the details). With $\epsilon > 0$ sufficiently small, we can define \emph{holonomy maps} $h_p$ (respectively, $h_q$) from $I_p$ (respectively, $I_q$) onto some interval along $\ell_\lambda$, $\tilde{I}_p$ (respectively, $\tilde{I}_q$), such that this map is a $C^{1 + \alpha}$ diffeomorphism and for every leaf $\eta\in W^s_\lambda$, $h_p$ (respectively, $h_q$) maps $\eta\cap I_p$ (respectively, $\eta\cap I_q$) onto $\eta\cap\tilde{I}_p$ (respectively, $\eta\cap\tilde{I}_q$); see Figure \ref{fig:holonomy} for an illustration. It follows that $\ell_\lambda\cap W^s_{\lambda}$ contains two Cantor subsets, $\tilde{I}_p\cap W^s_{\lambda}$ and $\tilde{I}_q\cap W^s_{\lambda}$, which can be realized as the attractors of two IFSs, one containing the function $h_p\circ f^{-n_p}\circ h_p^{-1}$, and the other $h_q\circ f^{-n_q}\circ h_q^{-1}$, having the fixed points $W^s(p)\cap\ell_\lambda$ and $W^s(q)\cap\ell_\lambda$, respectively.

Notice that for $s = p, q$,
\begin{align*}
-\log df^{-n_s}|_{W^u(s)}(s) = \log\norm{Df^{n_s}_s(E^u_s)},
\end{align*}
where $E^u(s)$ is a unit vector in the unstable subspace of $T_s S_\lambda$, and $Df^{n_s}_s$ is the differential of $f^{n_s}$ at the point $s$ (this follows from the fact that $f$ is area preserving). We call $\norm{Df^{n_s}_s(E^u_s)}$ \emph{the unstable multiplier} of $s$.

Let us now show that there exists a countable $D\subset (0, \infty)$, such that there exist periodic $p, q\in S_{\lambda}$ with
\begin{align}\label{eq:per}
\frac{\log\norm{Df^{n_p}_p(E^u_p)}}{\log\norm{Df^{n_q}_q(E^u_q)}}\notin\mathbb{Q}\hspace{2mm}\text{ for all }\hspace{2mm}\lambda\in(0,\infty)\setminus D.
\end{align}
(Dependence on $\lambda$ in \eqref{eq:per} is implicit in the notation; namely, $p, q\in S_\lambda$).

For $a\in (0, \infty)$, let us consider the two periodic points, $p_a$ and $q_a$ in $S_a$, given by
\begin{align*}
p_a = (-\frac{1}{2}, g_p(a), -\frac{1}{2})\hspace{2mm}\text{ and }\hspace{2mm} q_a=(0, g_q(a), 0),
\end{align*}
where
\begin{align*}
g_p(a) = \frac{1 + \sqrt{9 + 16a}}{4}\hspace{2mm}\text{ and }\hspace{2mm}g_q(a) = \sqrt{a+1}.
\end{align*}
Let us compute the unstable multipliers of $p_a$ and $q_a$, which are given by the larger (in absolute value) of the two roots of the following equations, respectively (see \cite[p. 850]{Roberts1994}).
\begin{align*}
\mu + \frac{1}{\mu} = 8g_p(a)(1-2g_p(a))+1\hspace{2mm}\text{ and }\hspace{2mm} \mu +\frac{1}{\mu}=2(8g_q(a)^4+1).
\end{align*}
After computing the larger (in absolute value) of the two roots for each equation, we obtain, respectively,
\begin{align}\label{eq:perp}
-\frac{8g_p(a)(1-2g_p(a))+1 - \sqrt{(8g_p(a)(1-2g_p(a))+1)^2-4}}{2}
\end{align}
and
\begin{align}\label{eq:perq}
8g_q(a)^4+1+\sqrt{(8g_q(a)^4+1)^2 - 1}.
\end{align}
Clearly both are analytic in $a$ for $a\in(0,\infty)$ and differ for values of $a$, so \eqref{eq:per} follows. But then it follows that the pair of iterated function systems determining the Cantor sets $\tilde{I}_p\cap W^s_\lambda$ and $\tilde{I}_q\cap W_\lambda^s$ satisfies the hypothesis of \cite[Corollary 1.5]{Hochman2012} (see the paragraph following the statement of the corollary). This gives
\begin{align*}
\shd_\lambda^2\geq \min\set{2\shd_\lambda, 1}
\end{align*}
(the inequality follows from the fact that the Hausdorff dimensions of $\tilde{I}_p\cap W^s$ and $\tilde{I}_q\cap W^s$ coincide with $\shd_\lambda$). On the other hand, we have \eqref{eq:ineq}.

The proof is finished by noting that $\ell_\lambda$ is a smooth embedding of $\R$ into $S_{\lambda}$. It is also clear that the same technique can be applied to prove Theorem \ref{thm:thm1}.

\section*{Acknowledgement}
I greatfully acknowledge helpful correspondence with David Damanik, Anton Gorodetski, Carlos Moreira, and Pablo Shmerkin. I would also like to thank Jake Fillman and Boris Solomyak for their helpful remarks.

\providecommand{\bysame}{\leavevmode\hbox to3em{\hrulefill}\thinspace}
\providecommand{\MR}{\relax\ifhmode\unskip\space\fi MR }
\providecommand{\MRhref}[2]{%
  \href{http://www.ams.org/mathscinet-getitem?mr=#1}{#2}
}
\providecommand{\href}[2]{#2}

\end{document}